\documentclass[conference]{IEEEtran}

\usepackage[utf8]{inputenc}
\usepackage{microtype}
\usepackage{graphicx}
\usepackage{amsmath}
\usepackage{bm} 
\usepackage{algorithm,listings}

\begin{document}

\title{Node-type-based load-balancing routing \\
       for Parallel Generalized Fat-Trees}

\author{\IEEEauthorblockN{John Gliksberg}
\IEEEauthorblockA{\textit{UVSQ,\ UCLM,\ Atos} \\
Versailles, France \\ john.gliksberg@uvsq.fr}
\and
\IEEEauthorblockN{Jean-Noël Quintin}
\IEEEauthorblockA{\textit{Atos} \\
Les Clayes-sous-Bois, France \\ jean-noel.quintin@atos.net}
\and
\IEEEauthorblockN{Pedro Javier García}
\IEEEauthorblockA{\textit{UCLM} \\
Albacete, Spain \\ pedrojavier.garcia@uclm.es}
}

\maketitle

\begin{abstract}
High-Performance Computing (HPC) clusters
are made up of a variety of node types
(usually compute, I/O, service, and GPGPU nodes)
and applications don't use nodes of a different type the same way.
Resulting communication patterns reflect organization of groups of nodes,
and current optimal routing algorithms for all-to-all patterns
will not always maximize performance for group-specific communications.
Since application communication patterns are rarely available beforehand,
we choose to rely on node types as a good guess for node usage.
We provide a description of node type heterogeneity
and analyse performance degradation caused by unlucky repartition
of nodes of the same type.
We provide an extension to routing algorithms
for Parallel Generalized Fat-Tree topologies (PGFTs)
which balances load amongst groups of nodes of the same type.
We show how it removes these performance issues by comparing results
in a variety of situations against corresponding classical algorithms.
\end{abstract}


\section*{Introduction}

Routing algorithms for HPC systems aim, for one thing,
to avoid congestion during application execution.
No perfect agnostic algorithm exists, and designing a good routing algorithm
usually requires paying attention to the topology
and communication patterns which will take place.
As detailed by Vigneras \& Quintin~\cite{vigneras2015fault},
we can consider that the topology of an HPC cluster never changes,
so algorithms are usually designed for a given topology class
(i.e.\ topology-aware algorithms).
On the other hand, communication patterns are hard to observe
(it is distributed information and can evolve rapidly),
rarely known in advance
(sometimes unpredictable, sometimes not predicted, sometimes classified),
and in the case of multiple applications running concurrently
it is potentially impossible to reroute the cluster on-the-fly
for optimal performance of each application without causing deadlocks;
indeed, there are few algorithms which rely on real communication patterns.
Existing research instead focuses on common worst case scenarios:
scatter, gather, n2pairs, all-to-all, hot-spots, etc.

We observe that when nodes are not all of the same type
(compute, I/O, service, GPGPU, …)
communication patterns for applications will usually be subsets
of worst-case scenarios with only one type of source nodes
and one type of destination nodes.
In the case of parallel generalized fat-trees (PGFTs),
it is intuitive to observe how existing load-balancing algorithms
(namely Dmodk and Smodk)
can result in avoidable congestion during type-specific communication phases.
The aim of this article is to propose new routing algorithms
which will provide the same performance for type-specific patterns
as existing ones do for type-unspecific patterns.

Section~\ref{sec:related} describes the existing context in detail
to improve understanding of the performance issues.
Characteristics of fat-tree topologies and their routing algorithms
are presented to ease analysis of routing algorithms' quality.
A case-study to this effect is introduced
alongside a description of node type heterogeneity
in Section~\ref{sec:heterogeneity}.
A corresponding communication pattern is chosen in Section~\ref{sec:analysis},
alongside a statically-computed congestion metric;
three routing algorithms are then analysed in detail
to show how they under-utilize available network resources.
Section~\ref{sec:gxmodk} provides a new technique to use these resources
more efficiently without losing properties of the existing algorithms.

\section{Background}%
\label{sec:related}

\subsection{Types of fat-tree topologies}

Fat-tree topologies were introduced by Leiserson~\cite{leiserson1985fat}
for their high capacity to represent any network for a given size.
All fat-tree topologies are deadlock-free when routed with shortest paths;\@
this property is one of the main advantages of fat-trees.
$K$-ary $n$-trees were subsequently formalized
by Petrini \& Vanneschi~\cite{petrini1997k}
and describe real-life implementations of fat-trees with low-radix switches
while being rearrangeably-non-blocking for any n2pair pattern.
Extended Generalized Fat-Trees (XGFTs),
introduced by Ohring et al.~\cite{ohring1995generalized},
describe a more general class of topologies which keep some properties
of $k$-ary $n$-trees but allow building much smaller and cheaper networks
with only partly reduced overall performance.
These topologies are generally not capable of providing
full cross-bisectional bandwidth (CBB) for a given
number of end-nodes and switch radix.

Zahavi introduced Parallel Generalized Fat-Trees
which can provide slimmed topologies with full CBB~\cite{zahavi2010d};
they are also useful for their fast tolerance to faults on duplicated links.
PGFTs are defined by their number of levels $h$,
the upwards arity $w$, downwards arity $m$, and link parallelism $p$,
each parameter for every level in the topology.
(The corresponding function is
$PGFT(h; m_0, \ldots, m_{n-1}; w_0, \ldots, w_{n-1}; p_0, \ldots, p_{n-1})$.)
They are built recursively with duplicated subgroups
composed of smaller PGFTs
(containing only interconnected switches of lower levels).
Each switch is addressed with a tuple $(l, a_h, \ldots, a_l)$
where l is its level
and $a$ is the vector describing the sub-trees the node is located at.

\subsection{Vocabulary}

Fat-trees are composed of levels
consisting of switches connected above and below.
As a result, elements related to a switch can be classified
using an \emph{up} or \emph{down} prefix.
For example \emph{up-switches} are switches linked to the switch in question
which are in the level above; the same logic applies to its ports and links.

\emph{Up} and \emph{down}-routes also relate to the direction level-wise.
It is worth pointing out that this differs
from existing topology-agnostic Up/Down routing algorithms,
where \emph{up} means ``towards the root node'' and conversely for \emph{down}.

\emph{Top}-switches are those in the highest level;
\emph{leaf}-switches are (as in all indirect topologies)
those connected to end-nodes.
For fat-trees we call all switches in the lowest level \emph{leaves}.

Hereafter, \emph{L1} switches are those at the first level (leaves),
followed by \emph{L2}s, etc.

\subsection{Side-note on adaptive routing}

Adaptive routing can react to congestion by diverting communication
towards alternative routes.
In the case of network congestion, spreading congested traffic
as much as possible is a right approach, because the congestion
was caused by unnecessarily colliding traffic flows in the first place.

Congestion can, however, originate from end-nodes themselves
(it can then be referred to as end-node congestion),
in that case spreading congested traffic will not solve the situation
and will instead further increase the problems arising from that traffic
to the rest of the topology.
In particular, more traffic flows will share paths
with congested traffic, hence increasing the probability
of the latter causing head-of-line blocking
to the former~\cite{rocher2017impact}.
Furthermore, traffic that was routed adaptively loses the property
of being transmitted in-order, potentially causing supplementary cost
to the application or communication layer;
for that reason the communication layer can mark packets
to forbid them from being routed adaptively.

Since adaptive routing cannot differentiate
between end-node congestion and network congestion,
it does not undermine the need for high-quality deterministic routing.
Instead, research focuses on techniques to either
reduce the potentially harmful collateral impact of adaptive routing
or reduce congestion from happening in the first place
with injection-throttling~\cite{escudero2011combining} or better deterministic routing.
The latter is the aim of this article.

\subsection{Overview of routing algorithms across fat-tree topologies}

Pure fat-trees are never used in supercomputers,
because they rely on very high-radix switches when there are many nodes.
However if we were to route them, that would simply mean
following the single shortest path available between any two nodes.
Any pair of nodes is associated with a single switch
at an equal and minimal distance from both,
it is called the nearest common ancestor (NCA).


When routing $k$-ary $n$-trees, every pair of nodes has multiple NCAs.
Optimal routing then comes down to distributing NCAs via which to route
to avoid network-congestion from happening in the first place.
For XGFTs and PGFTs, the problem is extended to take into account
per-level arities and parallel links.

\subsubsection{Random routing}

When multiple NCAs are available in fat-tree topologies,
one approach to balancing the load of deterministic routes
is to randomly choose upwards routes.
There is only one downward route from a switch to a node.
However in PGFT topologies there is a choice
among parallel links. This choice is made randomly too.

On average, the routes are randomly load-balanced:
all-to-all traffic will not cause implicit bias
towards any part of the topology.
Deviations from the average will, however, cause routes to overlap
and induce network congestion.

\subsubsection{Dmodk routing}

The regular structure of fat-tree topologies can be used
to uniformly distribute routes and achieve load-balancing routing
with a deterministic method based on packet destination ID,
instead of random routing.
Zahavi defines such a routing algorithm for PGFTs in a closed form
with upwards routes $P^U$ leading to destination $d$
for all switches in level $l$ computed as follows~\cite{zahavi2010d}:

\[
    P_l^U(d)
    := \left\lfloor \frac{d}{\prod_{k=1}^{l}w_k} \right\rfloor
    \bmod \left( w_{l+1} p_{l+1} \right)
\]

This formula assigns an index among the switch's \emph{up-ports},
which must be indexed beforehand to match the topological addressing scheme.
All switches in a level that are not in the same subgroup as the destination
are assigned the same upwards route.
(Those that \emph{are} in the subgroup must be routed downwards.)
This formula and its corresponding algorithm are called D-mod-k or Dmodk.
This method can be simplified for fat-trees simpler than PGFTs
and in all cases balances the load while concentrating routes
to the same destination, thus concentrating the undesired effects
of same-destination end-node congestion within a single-root subtree.

In the case of PGFTs, parallel links are indexed in a round-robin manner
so that all \emph{up-switches} are assigned a route
before multiple routes are assigned towards a single switch
(via those multiple links).
This, combined with the above formula, ensures a distribution similar
to the one defined sequentially in Zahavi et al.'s previous work
concerning fat-trees~\cite{zahavi2010optimized}.

Gomez et al.~\cite{gomez2007deterministic} routes $k$-ary $n$-trees
with a method which applies bitmasks to the destination number.
This method is defined in detail for $k=2$;
a similar approach can be extended for higher values of $k$.
This algorithm can be considered as a specialized version of Dmodk routing.

\subsubsection{Smodk routing}

If switches can determine the sources of messages,
routing algorithms can use that information as well.
From this an alternative to Dmodk can be defined:
Smodk, which propagates messages similarly to Dmodk
but based on source node ID rather than destination {ID}.
This algorithm concentrates together routes from the same source,
thus concentrating the undesired effects of same-source end-node congestion
as much as possible.

In cases where communications are symmetrical between patterns
with several destinations per source
and those with several sources per destination,
there is no reason for Dmodk or Smodk to be better than the other.
Otherwise there isn't necessarily one choice which is always better,
but choosing Smodk for multiple-destination heavy patterns
(and Dmodk for multiple source heavy patterns)
is a reasonable heuristic~\cite{rodriguez2009oblivious}.

We refer to Dmodk and Smodk together as a class of algorithms
as Xmodk for the rest of this article.

\section{Heterogeneous clusters}%
\label{sec:heterogeneity}

Supercomputers are often clusters made of several types of nodes,
rather than the common description of a single type of computing nodes.
Other types of nodes can include IO nodes for short and long-term data storage;
service or management nodes for login, node reservation, deployment,
monitoring, fault-tolerance;
GPGPU and FPGA nodes for optimized computations

There are various strategies to place secondary nodes (IO or service)
in existing clusters, which are usually not described in research material.
In the case of fat-trees, strategies can include:
\begin{itemize}
    \item Placing a constant number of secondary nodes of each type
        at every leaf
    \item Adding an irregular subgroup with secondary nodes
        connected to the top switches like the other regular subgroups
        (this generally breaks fat-tree properties)
    \item Connecting the cluster to an external topology via routers.
        For example if using a Lustre file system, Lustre routers can be nodes
        of the cluster leading to an array of IO servers
        of which the fabric management and routing algorithm are not aware.
\end{itemize}

As a concrete example, BXI%
\footnote{BXI is the interconnect technology developed by Bull/Atos.
It comprises hardware (switches, links)
and software (firmware, low-level and high-level development environment
on which are built the fabric management and routing algorithms).}
switches have 48 ports.
Some BXI switch have only copper ports,
some others have three optical ports.
The optical ports are placed identically on all switches
and are dedicated to nodes physically far within the topology
(i.e.\ management nodes and IO proxy nodes).

\section{Analysis of a type-based \\ communication pattern}%
\label{sec:analysis}

\begin{figure}[h]
    \centering
    \includegraphics[angle=-90,width=\linewidth]{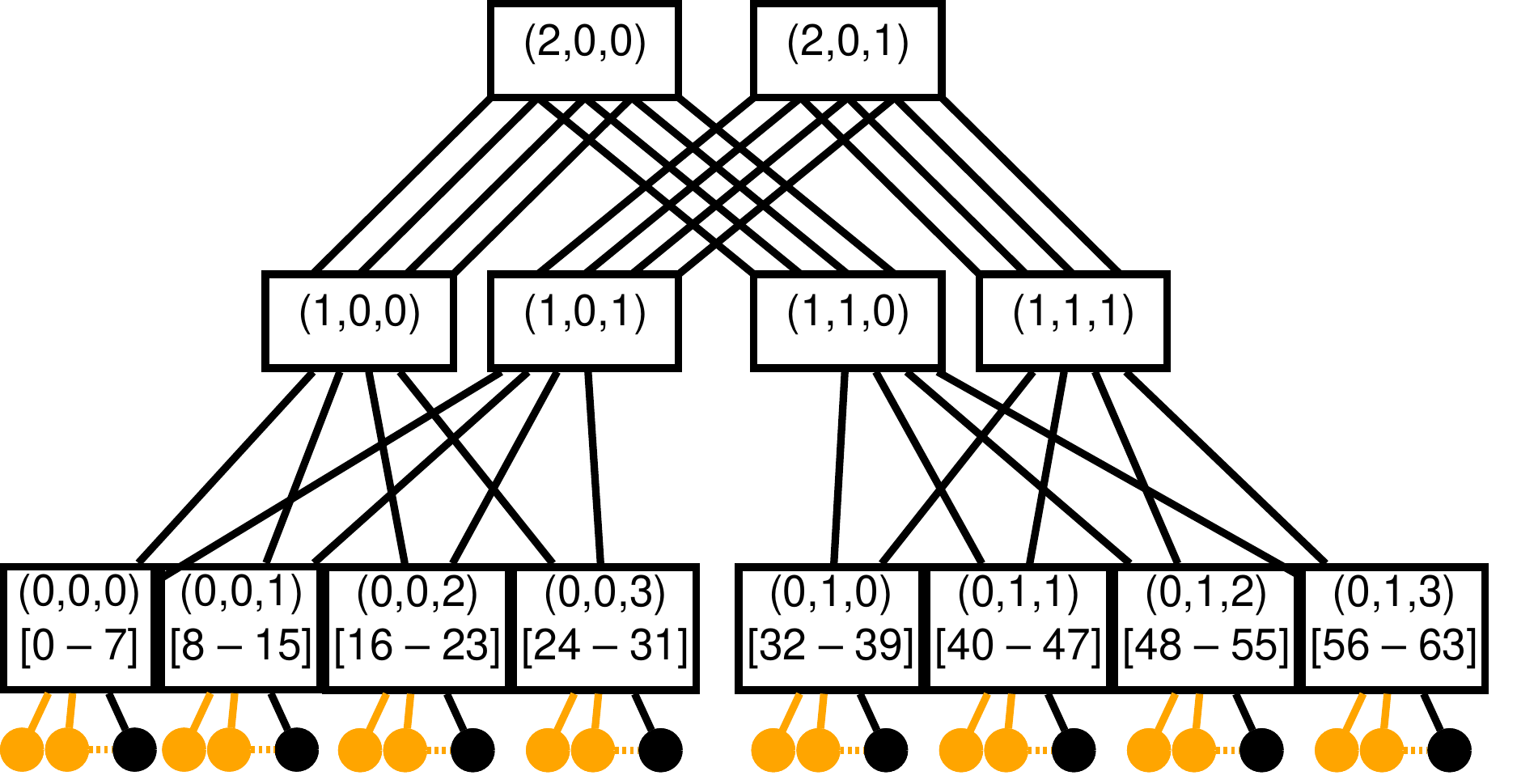}
    \caption{Case-study topology
        $PGFT(3; 8,4,2; 1,2,1; 1,1,4)$ —
        IO nodes (in black) have the largest NID of every leaf
        ($7, 15, 23, \cdots \equiv 7 \bmod 8$)}%
\label{fig:studypgft}
\end{figure}

We will use a simplified case-study to show
how node-type-oblivious load-balancing routing
can result in unnecessary network congestion.
The topology for this case-study is a pruned 3-level PGFT
with low-radix switches and nonfull cross-bisectional bandwidth (CBB)
(see figure~\ref{fig:studypgft}).
Nodes are indexed by port rank on their leaf
and by leaf address comparison between leaves.
The last port of every leaf is reserved for IO nodes;
they have NIDs whose modulo by 8 is 7.
We use a topology with nonfull CBB because otherwise
there would be no possible congestion at any top-port.

This case-study is based on a communication pattern
commonly found in distributed applications:
data collection from all compute nodes to IO nodes,
each compute node sending to the IO node of its symmetrical leaf
(e.g.: $(0,0,1)$ is symmetrical to $(0,1,1)$,
so NIDs 8 to 14 send to NID 47).
In this case all routes will have to go through a top-switch.
This might be contained in a short time frame, following a barrier,
or spread out through the application lifetime, it does not matter.
For a given complete set of routes $R$,
we call $C2IO(R)$ the subset of routes affected by this pattern.

\newcommand{\MSD}[1]{\boldsymbol{C}_{#1}}
\subsection{Static congestion metric}

This study relies on a static metric to describe
the potential sources of contention.
The aim of this metric is to give a formal way to describe contention,
which abstracts the fine-grain causes of latency
to help build a general understanding of how to avoid contention.
This contrasts with common techniques based on simulation
or experimentation which do not link observations of contention
with a corresponding explanation.
This simple technique of estimation of contention is new;
it is also used in concert with the architecture described by
Vigneras \& Quintin~\cite{vigneras2015fault} with the goal of automating
computation of that metric for potential integration
into the fabric management's decision making.
Analysis in terms of this metric is sufficient
to prove and explain drawbacks and benefits of algorithms,
but a simulation-based analysis would complement this work
to give tangible results for real-life applications.

Worst-case scenario contention can be measured by the number
of possible flows going through a port at the same time.
Once the topology is routed, if a given port $p$ is used as output for routes,
we can count the number of distinct sources ($src(R,p)$)
and destinations ($dst(R,p)$) for these routes:

\begin{itemize}
    \item Both values are non-nil, since there are routes
        going through the port.
    \item If one of these values is equal to 1, this port will never be used
        for unrelated communications;\@
        the port is subjected to only one flow of communication.
        Any potential packet concurrency at the port means that
        there is a corresponding concurrency at the sending end-node
        or receiving end-node.
        That end-node will be the cause of unavoidable congestion
        of an order of magnitude more important,
        and no packet from another flow could be affected.
        \begin{figure}[h!]
            \centering
            \includegraphics[angle=-90,width=.5\linewidth]{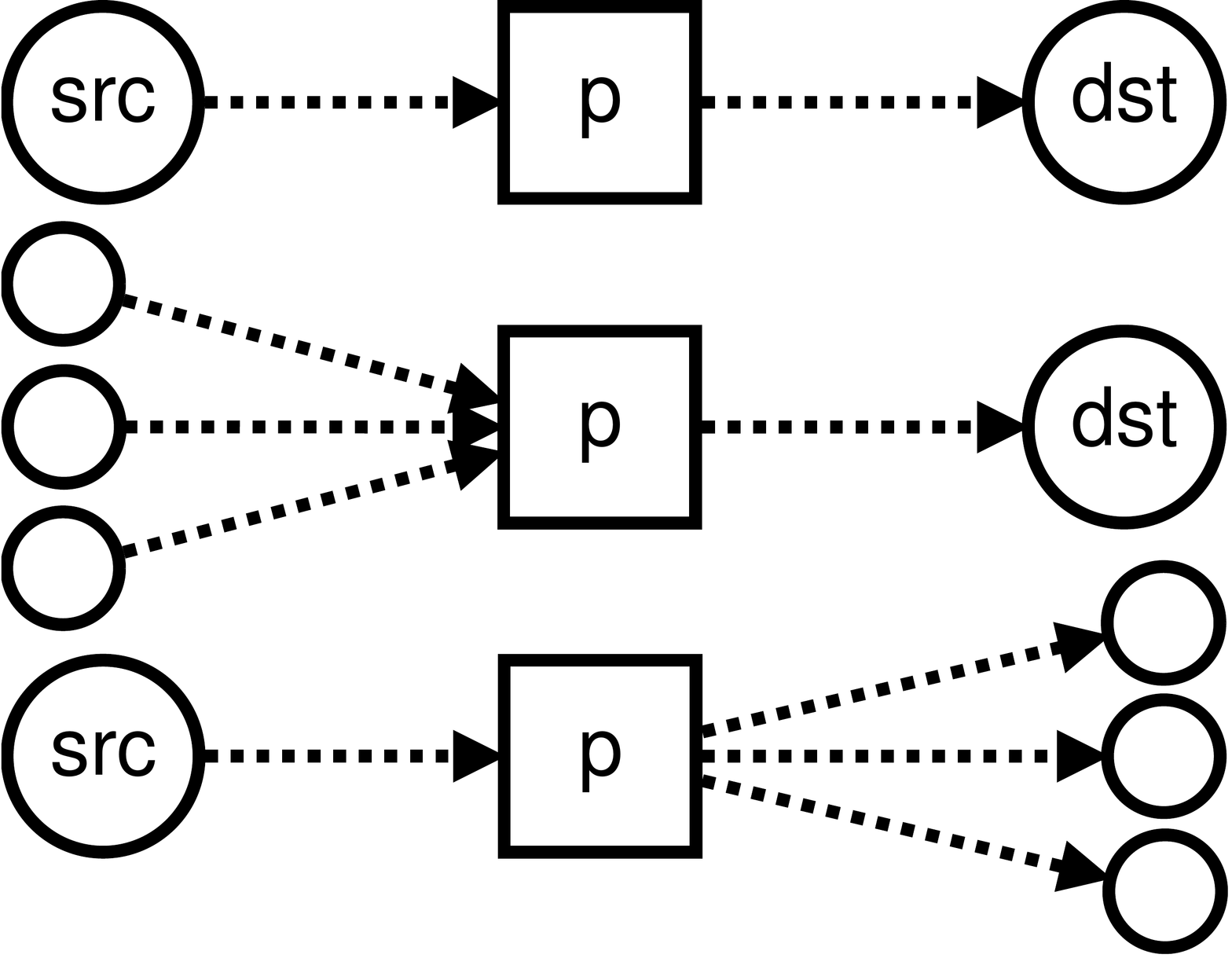}
            \caption{Example sets of routes for which $p$
                     is subjected only to one flow}\label{cong_min}
        \end{figure}
    \item If both values are greater than one, there are unrelated communications
        that might interact at this port.
        This can lead to potentially avoidable network congestion.
        \begin{figure}[h!]
            \centering
            \includegraphics[angle=-90,width=.3\linewidth]{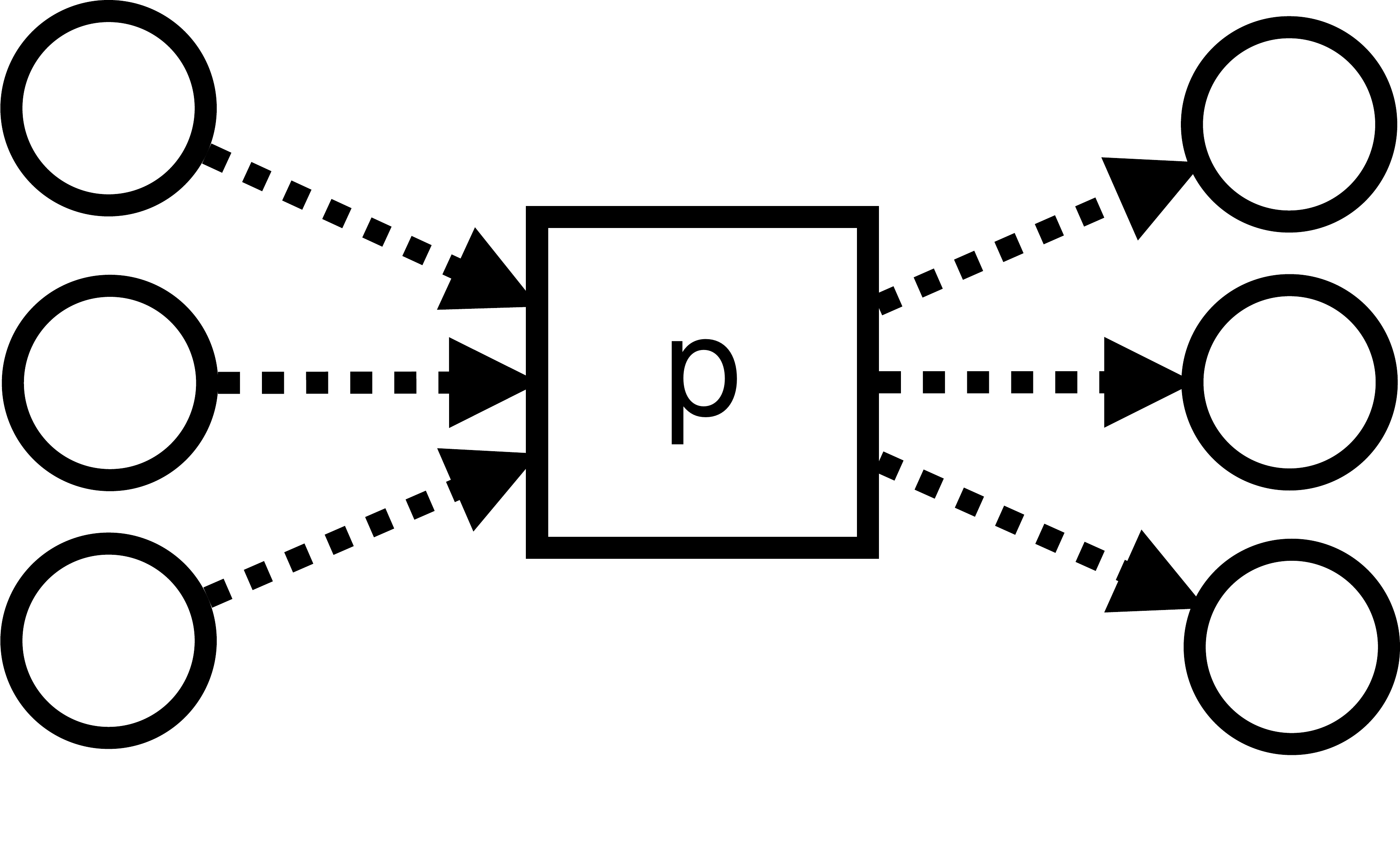}
            \caption{Example routes for which $p$
                     can be subjected to multiple flows}\label{cong_nonmin}
        \end{figure}
\end{itemize}

For a given set of routes R, we call this metric $\MSD{port}$:

\[
    \MSD{p}(R) := \min(src(R, p), dst(R, p))
\]

This metric does not imply there will always be network congestion
at ports with $\MSD{port} > 1$, but it shows the worst case.
Assuming all flows are similar,
collisions will happen more frequently when more flows are involved.
As a result we claim that in general port $a$
will tend to be more congested than port $b$
if $\MSD{a} > \MSD{b}$, even if it depends
on the exact timing of communications.
From this we can deduce a reasonable metric for the whole topology:

\newcommand{\MMSD}{\boldsymbol{C}_{topo}}
\[
    \MMSD(R) := \max_{p\in topo}(\MSD{p}(R))
\]

Routing in a balanced manner means minimizing that metric.
Studying a communication pattern means applying this metric
only to the routes affected by the pattern
rather than all the routes computed.

For this metric we consider ports as output for the routes,
but the same analysis can be made with ports as input.
This does not cause $\MMSD(R)$ to vary
when the pattern has symmetrical communications
between sources and destinations.


\subsection{Dmodk performance}

With Dmodk routing, destinations will be assigned one switch
through which to route in every subgroup.
More specifically, we will describe which up-port is routed as output
for switches not directly above the destination,
and which down-port is routed as output for switches directly above,
when there are several parallel ports available.

$w_1$ = 2, $p_1$ = 1: every destination is assigned the L2 switches
corresponding to its index modulo two.
(E.g.: $47\bmod2=1$,
thus destination 47 is assigned the second L2 switch of each subgroup.)
The eight IO destinations are all assigned the same two L2 switches
($(1,0,1)$ and $(1,1,1)$),
and more specifically the last up-port of the L2 switch
not in their subgroup.

$w_2 = 1$: there is only one L3 switch each L2 switch leads to.
It still corresponds to the destination's index modulo 2.
IO destinations are assigned the second L3 switch.

\begin{figure}[h]
    \centering
    \includegraphics[angle=-90,width=\linewidth]{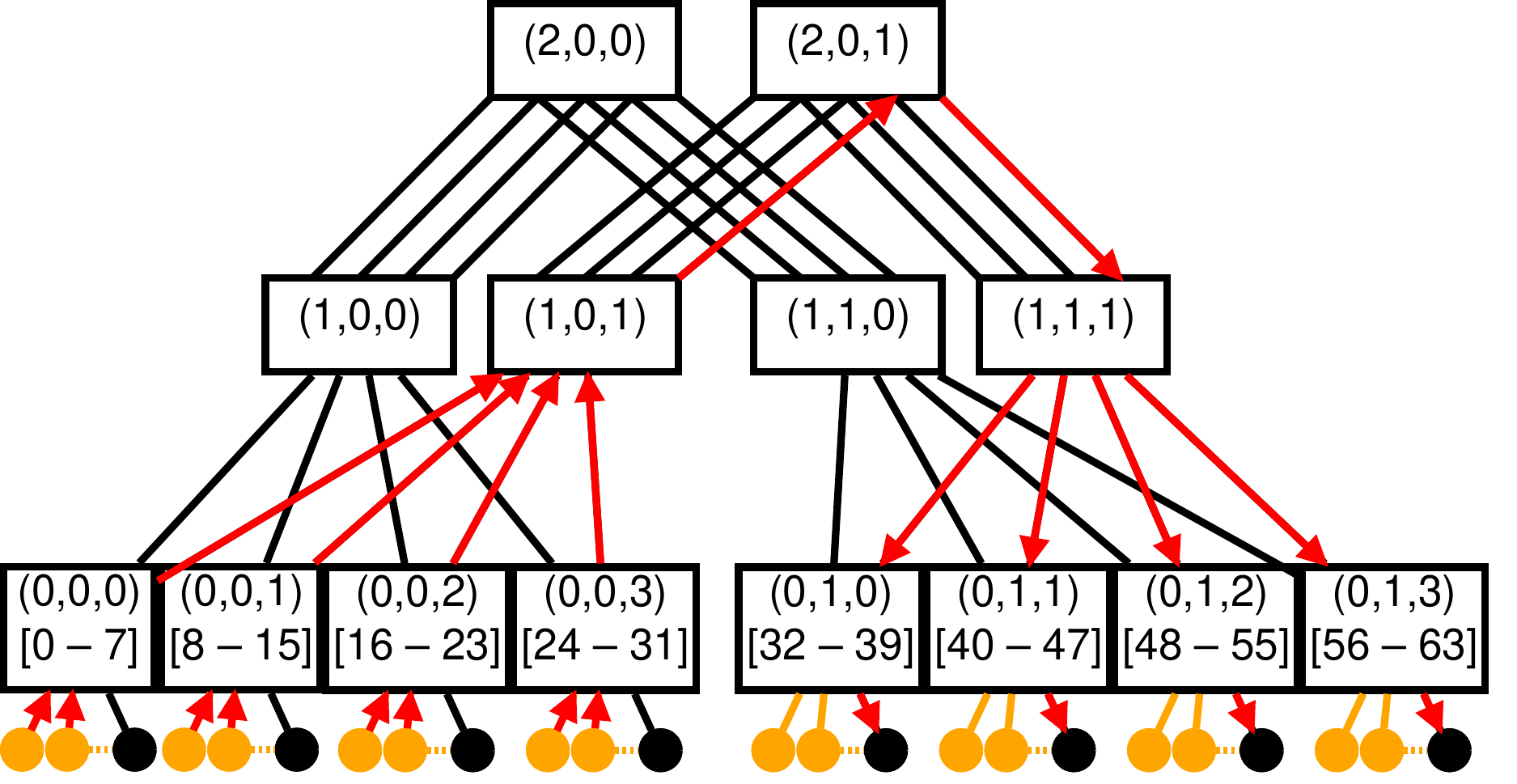}
    \caption{Set of all routes (in red) going towards IO nodes
        of the right subgroup, under Dmodk routing.
        $(2,0,1)$'s port with highest rank is used as output for all routes.}%
\label{fig:studypgft2_dmodk}
\end{figure}

$p_2 = 4$: they are more specifically assigned
the last port of the four leading to their subgroup.
This leaves four destinations per top-port.
Figure~\ref{fig:studypgft2_dmodk} shows this for $(2,0,1)$.

Furthermore, each destination has exactly one corresponding source:

\begin{align*}
    \MSD{(2,0,1):7}(C2IO(Dmodk)) &= \MSD{(2,0,1):8}(C2IO(Dmodk)) \\
                                 &= \min(56,4) = 4
\end{align*}

$(2, 0, 1):7$ is the last of $(2, 0, 1)$'s four ports
leading to the left subgroup,
and $(2, 0, 1):8$ is the last leading
to the right subgroup.

There are 8 (leaves) times 7 (compute nodes per leaf)
=~56 compute destinations,
to which are assigned all top-ports except for the two
ports of $(2,0,1)$ assigned to IO nodes
=~14 top-ports, in a balanced manner;
this leaves four compute destinations per port.
None of these routes are affected by $C2IO$, therefore,

\begin{align*}
    \forall p \notin (2, 0, 1),\quad \MSD{p}(C2IO(Dmodk)) &= 0 \\
    \MMSD(C2IO(Dmodk)) &= 4
\end{align*}

To reformulate this result:\ for the given communication pattern
most of the top-ports are unused
while two top-ports have a strong risk of congestion.
This can be seen for one of these ports on Figure~\ref{fig:studypgft2_dmodk}:
\ the routes shown by the red arrows concentrate around the top-port
like those shown in Figure~\ref{cong_nonmin}.

This is sub-optimal, while spreading both subgroups of four IO destinations
any disjoint way among the 8 ports leading to each in the top-switches
would have lead to $\MMSD(C2IO(R_{dst})) = 1$.
The object of Section~\ref{sec:gxmodk} will be to define
such a set of routes $R_{dst}$.

\subsection{Smodk performance}

With Smodk, routes from compute to IO nodes are spread per source.
With the same process as Dmodk for compute nodes as destinations,
we determine which ports are used with Smodk for computed nodes as sources.
More specifically, we will describe which down-port is used as output
for switches not directly above the source,
and which up-port is used as output for switches directly above,
when there are several parallel ports available.

\begin{figure}[h]
    \centering
    \includegraphics[angle=-90,width=\linewidth]{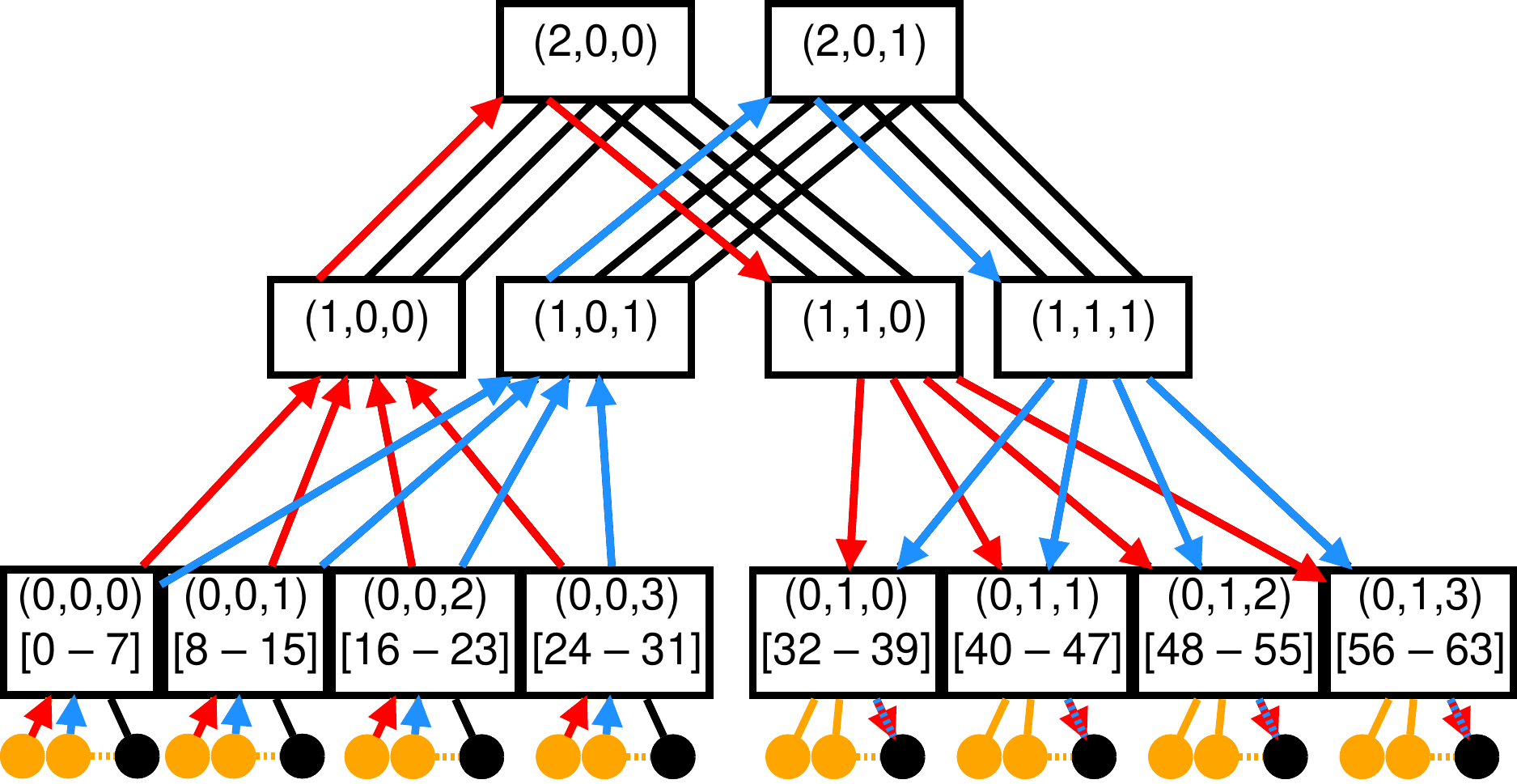}
    \caption{Two subsets of $C2IO(Smodk)$:\@
        red routes have source NID $0 \bmod 8$,
        light blue routes have source NID $1 \bmod 8$.}%
\label{fig:studypgft2_smodk}
\end{figure}

There are 8 top-ports that lead to each subgroup,
and 28 compute sources per subgroup;
after every group of 7 sources, one NID is skipped,
which corresponds to skipping the last considered port of $(2,0,1)$.
We conclude that two ports of $(2,0,1)$ have no compute source,
and every other top-port has four compute sources
which are all connected to different leaves
and as a result send to different IO destinations.
This results in $\MSD{p}(C2IO(Smodk)) = 4$ for each affected top-port $p$.
This is shown in Figure~\ref{fig:studypgft2_smodk}.

This means that in this case there are fourteen top-ports
with a high risk of congestion;
The sets of routes depicted by red and light-blue arrows
in Figure~\ref{fig:studypgft2_smodk} also correspond to the situation
shown in Figure~\ref{cong_nonmin} to show how high congestion risk
for two of the fourteen concerned top-ports.
For this communication pattern Smodk is less suited than Dmodk.

Optimizing a source-based routing for this pattern and metric
means coalescing routes to the same destination.
This would be possible for a given pattern;
however this article aims to route based on node-type only,
therefore we cannot use specific distribution information.
If we always have colliding routes lead to distinct destinations,
the best we can reach in this situation is still $\MMSD(C2IO(R_{src})) = 4$.

Since the pattern considered has few destinations and many sources,
it is reasonable that a routing algorithm which concentrates routes
from the same sources will be difficult to improve.
The opposite will happen for the opposite pattern.
Section~\ref{sec:gxmodk} will also define the set of routes $R_{src}$.

\subsection{Random routing performance}

Dmodk was unable to reach $\MMSD(Dmodk) = 1$ because the modulo operation
depends on NIDs and has no information about the communication pattern.
Random routing does not depend on NID;\@
it spreads every route uniformly over the available ports,
and as a result every subset of routes is also spread uniformly.
Therefore $C2IO(Random)$ does not have particularly coalesced routes.

In practice, distributing each group of 28 routes
into its corresponding 8 top-ports
always causes collisions between routes that have different destinations.
The probability of collision is very close to 1.%
\footnote{Determining with what probability there will be a conflict
between two of the 28 routes (leading to different destinations)
spread through the 8 top-ports is an example of collision probability
between sets of random variables~\cite{wendl2003collision}
(a generalization of the girl/boy birthday problem).
However in this case the total number of variables
is greater than the number of choices.
In that situation, the article's formula for generalized number of sets
has a term which always cancels out for part of the computation;
it seemed possibly ill-adapted and was therefore discarded.}
Therefore, we can safely state that $\MMSD(C2IO(Random))$
is always greater than 1.
Repeated computation of Random routing
for the given topology and communication pattern
resulted in $\MMSD(C2IO(Random))$ values of either 3 or 4:
\ i.e.\ rarely better than Dmodk.

Random routing will usually give slightly better results
than Dmodk or Smodk as soon as the communication patterns
have a given bias, but they will always leave some ports
with avoidable congestion.
Just as Xmodk algorithms aim to compute perfect routes
for the general worst-case scenario,
we want to compute perfect routes
for the type-specific worst-case scenario.

\section{Grouped Xmodk}\label{sec:gxmodk}

In the previous section we show that the existing routing algorithms
do not balance the load correctly when the topology has mixed node types.

To improve routing for type-specific communication patterns,
we can use knowledge of node types and modify Xmodk algorithms.
The aim is to optimize resource usage depending on node type.
For example the optimization should achieve the best throughput
for communications towards IO proxies or compute nodes.

We suggest balancing each group of nodes separately
to improve load-balancing under worst-case type-specific patterns.
This corresponds to the previously mentioned $R_{dst}$ and $R_{src}$.

\subsection{Description of indexing}

Grouped Xmodk algorithms, or Gxmodk, consist of preprocessing NIDs.
Knowing each node's type, the algorithms begin by updating the NIDs accordingly,
as shown in algorithm~\ref{alg:gxmodk_preprocess}.

\begin{algorithm}
\begin{lstlisting}[language=python,tabsize=4,basicstyle=\footnotesize]
node_types = set((node.type for node in topo.nodes))
newnodes = list()
for node_type in node_types:
	for nid, node in enumerate(topo.nodes):
		if node.type == node_type:
			newnodes.append(node)
topo.nodes = newnodes
\end{lstlisting}
\caption{Reindex NIDs by type}\label{alg:gxmodk_preprocess}
\end{algorithm}

Re-indexing in the order of the original NIDs
ensures that consecutive reindexed NIDs are topologically close.

Xmodk is then applied as usual but with the updated NIDs.

\subsection{Analysis for previous topology and communication pattern}

We choose to call gNID a reindexed {NID}.
Let's suppose that compute nodes are reindexed first:
there are 56 so they are assigned gNIDs 0 to 55.
IO nodes are assigned gNIDs 56 to 63.
Now routing depends on whether Gdmodk or Gsmodk is used.

\subsubsection{Gdmodk results}

For Gdmodk, each IO destination is assigned a unique L2 switch
in each subgroup (e.g.:\ gNID 61 is assigned $(1,0,1)$ and $(1,1,1)$).
Each L2 switch is assigned two IO destinations of the opposite subgroup,
therefore the up-routes from L2 switches use only half
of the available parallel ports in a balanced manner.

\[ \MSD{p\in(1,*,*)}(C2IO(Gdmodk)) \le 1 \]

\begin{figure}[h]
    \centering
    \includegraphics[angle=-90,width=\linewidth]{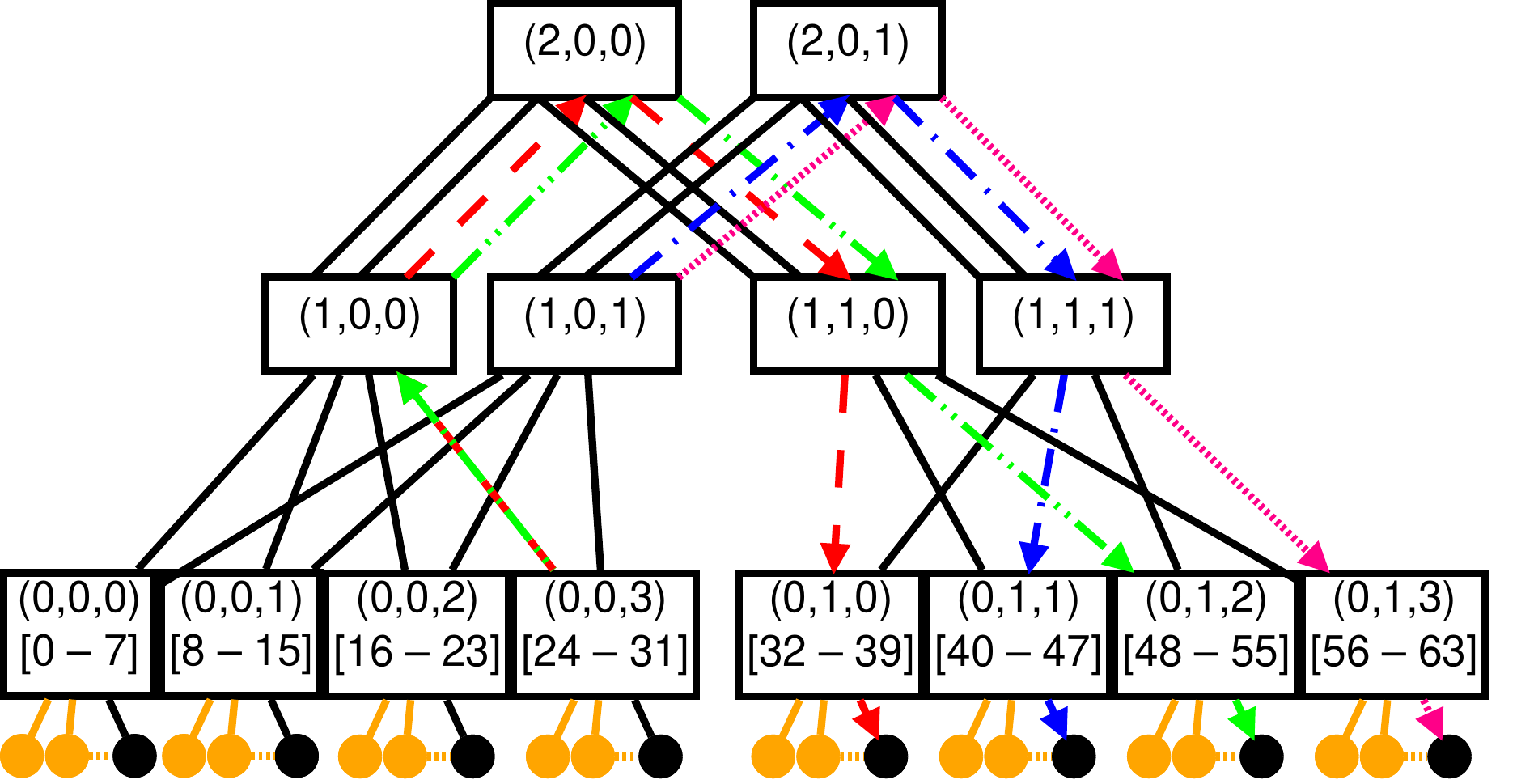}
    \caption{Simplified representation of all routes going to IO nodes
             in the right subgroup, under Gdmodk routing.
             Arrows not displayed can be deduced by shortest paths
             to displayed ones.}%
\label{fig:studypgft2_gdmodk}
\end{figure}

Each L3 switch is shared by two IO destinations for each subgroup
(e.g.: $(2,0,1)$ is shared by NIDs 15, 31, 47 and 63;
or gNIDs 57, 59, 61 and 63), which are assigned distinct output ports.
Figure~\ref{fig:studypgft2_gdmodk} shows how Gdmok distributes routes
efficiently when considering type-based communication patterns.
It can be interpreted by pointing out that each set of routes
specified by a given color has only one destination
(and matches the situation described in Figure~\ref{cong_min}),
with no overlap on output ports in the two upper levels
between two sets of routes.

\[ \MSD{p\in(\{1,2\},*,*)}(C2IO(Gdmodk)) = 1 \]

All leaves' up-ports have seven sources and two destinations.
$ \MSD{p\in Up-ports((0,*,*))}(C2IO(Gdmodk)) = 2$
as is shown with the overlapping dashed red and double-dotted green arrows
in Figure~\ref{fig:studypgft2_gdmodk}.
It is unavoidable for some of them
to have more than one for the given pattern,
so Gdmodk gives the best possible quality of routing tables.

\[ \MMSD(C2IO(Gdmodk)) = 2 \]

\subsubsection{Gsmodk results}

For Gsmodk, the 28 compute sources of each group are assigned
all 8 up-ports of the two L2 switches of their subgroup
in a balanced manner:\@ there are 7 compute sources per up-port
which are used to lead to 4 distinct IO destinations.
The 7 top-ports leading to the other subgroup are used the same way.

\[ \MMSD(C2IO(Gsmodk)) = 4 \]

\begin{figure}[h]
    \centering
    \includegraphics[angle=-90,width=\linewidth]{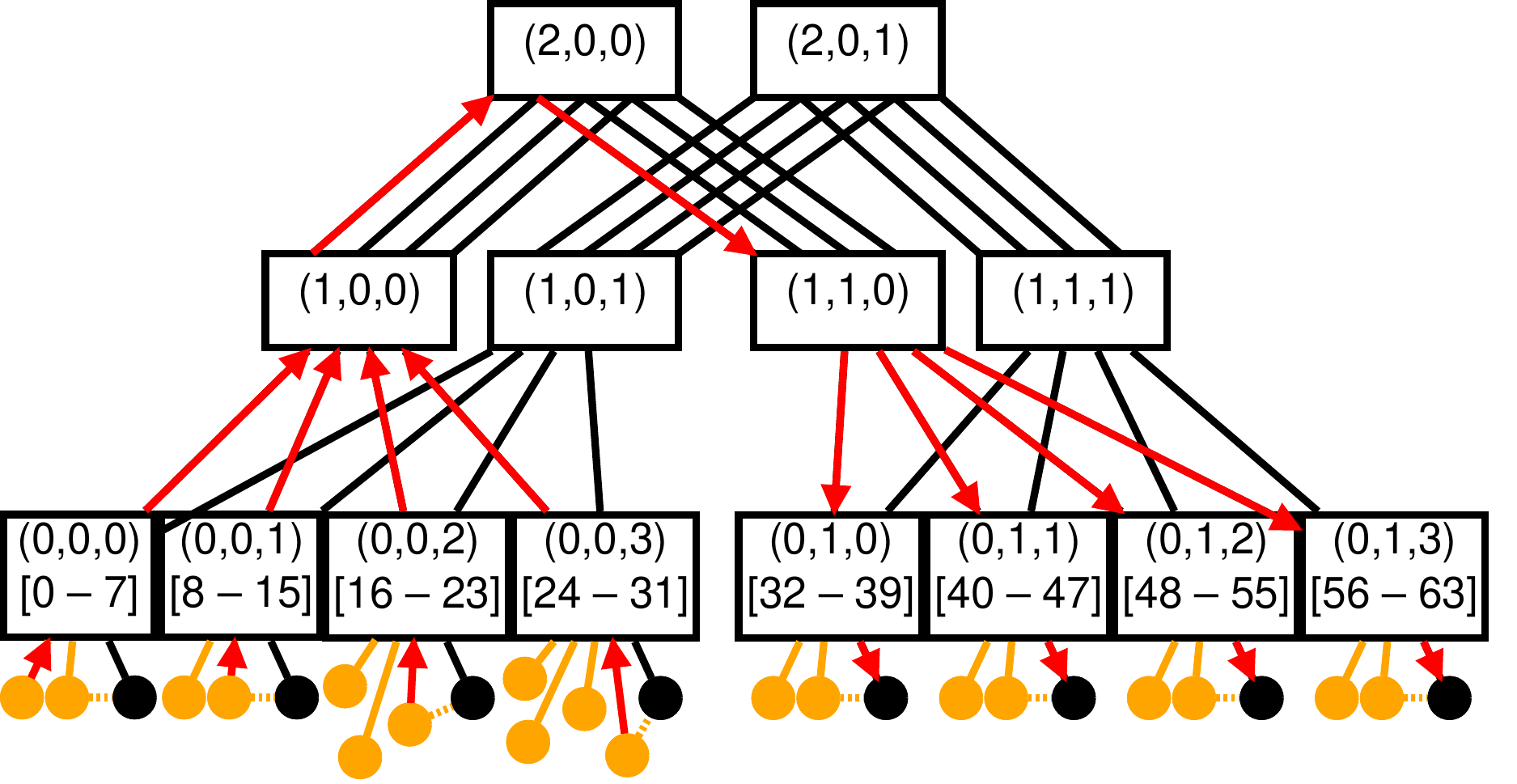}
    \caption{Subset of $C2IO(Gsmodk)$ with source gNID $0 \bmod 8$}%
\label{fig:studypgft2_gsmodk}
\end{figure}

Figure~\ref{fig:studypgft2_gsmodk} shows all routes of $C2IO(Gsmodk)$
that use one example top-port as output.

Gsmodk improves route distribution for this pattern compared with Smodk:\@
Since an eighth up-port is now used in both L2 switches $(1,*,1)$,
(and two down-ports of $(2,0,1)$),
each port now has 7 sources.
All of these ports that were used by Smodk had 8 sources.
This improvement is comparatively minor,
because few resources had been spared by Smodk on this pattern.
This shows that type-awareness doesn't solve the existing asymmetry issue
between optimizing routing to coalesce sources or destinations,
but it does improve routing with regards to type-specific patterns
any time Xmodk missed out on available resources.

On the symmetrical communication pattern
we would see the same improvement as we do between Dmodk and Gdmodk
for the considered communication pattern.
In general, if pattern $P$ is symmetrical to $Q$, we should always find:

\begin{align*}
    \MMSD(P(Dmodk)) &= \MMSD(Q(Smodk)) \\
    \MMSD(Q(Dmodk)) &= \MMSD(P(Smodk)) \\
    \MMSD(P(Gdmodk)) &= \MMSD(Q(Gsmodk)) \\
    \MMSD(Q(Gdmodk)) &= \MMSD(P(Gsmodk))
\end{align*}

\section*{Conclusions and future works}

In this paper we have defined realistic communication patterns
depending on node type which are present on our production cluster.
From this real-life scenario we analyzed
how existing solutions fare against these patterns.
We have specified how type-based communications
can result in unnecessary congestion.
To counter this we have provided new algorithms
to improve existing solutions.
We have shown a realistic example with, in one case,
a sevenfold decrease in congestion risk.

The congestion issue of Xmodk stems from nodes of a same type
having the same NID, modulo arities.
This also affects communications unrelated to node-type,
but optimizing for these means knowing about application usage.
Gxmodk aims only to improve the situation when node-type is known;
having early knowledge of applications' communication matrices
would warrant writing specific deterministic algorithms.

This article relies on a static flow metric
from which we deduce probable congestion.
A more thorough analysis of the relationship
between this metric and actual congestion
depending on fine-grain communication interaction
would also be warranted.
A corresponding study of the new algorithms
based on simulation rather than only a static congestion metric
would also provide results in terms of performance.

This work focuses on fat-trees, for which node indexing
allows intuitive understanding of NCA distribution;
from this we derive type-based pattern analysis
and devise a new node indexing to solve corresponding issues.
For other topologies (e.g.\ DragonFly, Generalized HyperCubes)
a similar work could also be attempted.
Furthermore, a procedural routing algorithm for fat-trees
(which can be useful for routing degraded fat-trees or similar topologies)
was omitted; a similar technique could be used to improve it.

\section*{Acknowledgements}

This BXI development has been undertaken under a cooperation
between CEA and Atos.
The goal of this cooperation is to co-design extreme computing solutions.
Atos thanks CEA for all their inputs
that were very valuable for this research.

This research was partly funded by a grant
of Programme des Investissments d’Avenir.
This work has been jointly supported by the Spanish MINECO
and European Commission (FEDER funds)
under the project TIN2015{-}66972{-}C5{-}2{-}R (MINECO/FEDER).

BXI development was also part of ELCI,
the French FSN (Fond pour la Société Numérique) cooperative project
that associates academic and industrial partners
to design and provide software components
for new generations of HPC datacenters.

\newpage
\bibliographystyle{plain}
\bibliography{routing}

\end{document}